# Band Oscillation Correlated with $T_c$ for Superconductors

Zhigang Song, Xin Li*

*John A. Paulson School of Engineering and Applied Sciences, Harvard University, Cambridge, MA 02138, USA*

**Abstract**

Previous theoretical studies on superconductivity were focused on the static states and adiabatic processes. Quantum mechanics simulations of time-dependent processes in superconductors were rarely performed previously. Here we use real-time time-dependent density functional theory to show a new phenomenon that the electron-phonon excitation is coherent and mode-selective, resulting in a periodic band oscillation. Surprisingly, the predicted oscillation frequency of charges or bands here are strongly correlated with a broad range of superconductivity transition temperatures of various main families of superconductors, including cuprates, Fe-based, hydrides, and $WTe_2$. This shed light on the pairing mechanism in unconventional superconductors, paving the way for the design of higher temperature superconductors.

lixin@seas.harvard.edu

Since the discovery of high-temperature superconductivity, the underlying mechanism has been elusive.[1] Although the conventional superconductivity can be explained by BCS theory, the prediction of the superconductive transition temperature ($T_c$) of different unconventional superconductors is still a challenge. Density functional theory (DFT) is one powerful tool to investigate microscopic mechanism of condensed matter materials. In previous works, the critical temperature $T_c$ is computed and investigated by the strength of electron-phonon coupling[2] for conventional superconductors. However, this method fails for unconventional superconductors. Some previous works on DFT find certain correlation between $T_c$ in unconventional superconductors and some materials-dependent factors. All these works, however, tried to predict high-$T$c superconductors based on static DFT calculation methods by solving the static equations, while the electron dynamics, nonadiabatic excitation, and electron coherence that are important to unconventional superconductivity are hard to include in early works.[3]

Recently, experiments show that ultrafast dynamics and coherent phonons are critical to the superconducting transition.[4-9] Some experiments report that light can induce, enhance, or switch the superconductivity because certain modes of coherent electron-phonon coupling can be enhanced by laser polarization.[4, 10-17] For example, light can induce superconducting-like behavior in $YBa_2Cu_3O_{6+\delta}$, and mid-infrared femtosecond pulses induce superconductivity phase transition in cuprates of $La_{1.675}Eu_{0.2}Sr_{0.125}CuO_4$.[15] Besides, photo-induced meltable states possibly competes with superconductivity.[18] In theory, the previous time-dependent simulations of semiconductors show phonons are coherent and can induce charge transfer, carrier scattering, and ultrafast structural transition in semiconductors.[19-24] Both experiments and simulations imply that the ultrafast dynamics is critical to the formation of superconductivity gap and electron pairing.

A pioneering theory of time-dependent Bogoliubov-de Gennes equations including exchange-correlation effects for superconductors was established by O.-J. Wacker, R. Kummel, and E.K. U. Gross in 1994.[25] Until now, however, it lacks a direct theoretical investigation on the ultrafast electron dynamics in real materials of superconductors, especially at DFT level. The nonadiabatic charge excitation, coherent electron phonon coupling, and other nonthermal processes require a time-dependent simulation.[26-31] Therefore, time-dependent first-principles study of ultrafast physics in femtosecond time regime becomes urgent, and more specifically for electron excitation by coherent phonons in superconductors.

In previous reports, real-time time-dependent density functional theory (rt-TDDFT) [32, 33] has shown to be a powerful tool to simulate the ultrafast structural phase transition.[34-38] In this letter, we performed rt-TDDFT simulation to study the effect of ultrafast electron dynamics in superconductors. The calculation is performed using the PWmat package [39-42] with SG15 norm conserving pseudopotential, PBE exchange correlation functional, and a plane wave energy cutoff of 50 Ry. In the rt-TDDFT calculation, the effective time step is 0.1 fs. A $k$-mesh of $3\times3\times1$ are used. The NVE ensemble is used for nuclear dynamics, where the atomic number, volume, and energy in the system are constants.

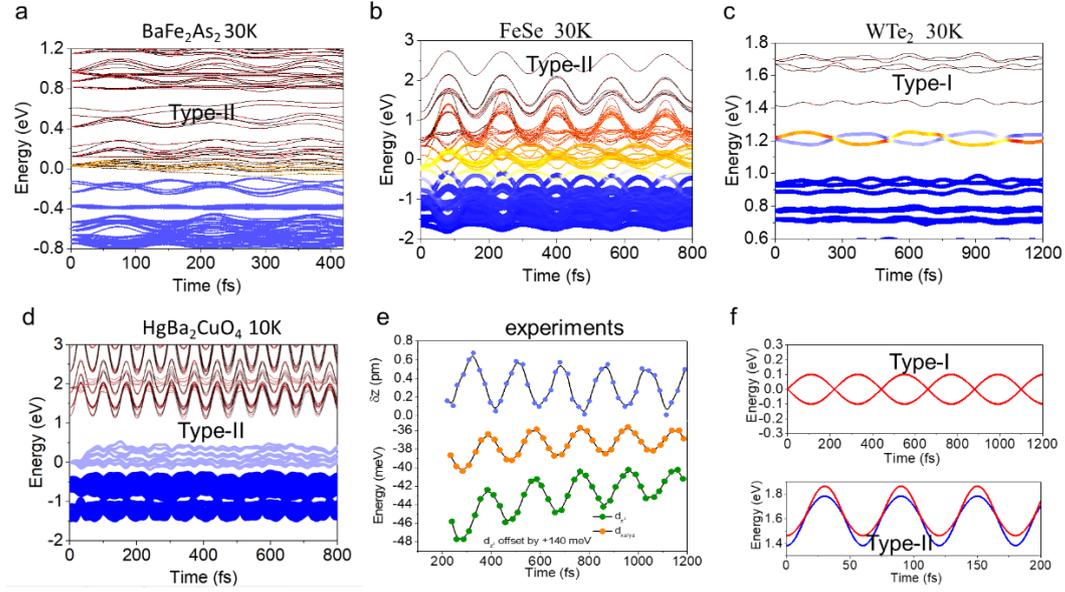

**Figure 1** (a-d) Energy at Γ point as a function of time in monolayer $WTe_2$, and bulk $BaFe_2As_2$, FeSe, and $HgBa_2CuO_4$. Both the weight and color represent the electron occupation. (e) The orbital-resolved energy shift form ultrafast x-ray experiments of FeSe.[43] (f) Energy shift obtained from our theoretical model. Upper and lower panels are Type-I and Type-II electron phonon couplings, respectively.

To study electron dynamics at a certain temperature, we performed rt-TDDFT calculations for different materials. The electronic structures are obtained by solving time-dependent Schrodinger equations considering as many phonons as possible. Fig. 1 shows the simulated instantaneous band energy. In the range of several hundreds of femtoseconds, we found a regulating oscillation of band structure for all the superconductors below the corresponding $T_c$. The instantaneous band edges as functions of time can be approximated by sine (cosine) curves. For example, the band oscillation period is 480 fs for a $WTe_2$ monolayer, where the bands are regularly opened and closed, which we call the type-I oscillation with such an out-of-phase pattern. The largest opening of bands due to such oscillation (e.g., the two bands at 1.2 eV in Fig. 1c) can reach about 0.1 eV at the temperature of 30K. Previously, a band oscillation with 420 fs was observed in the ultrafast ARPES experiment.[44, 45]

For other cuprates and iron-based superconductors, such as FeSe, $BaFe_2As_2$, $HgBa_2CuO_4$, $Hg_2BaCaCu_2O_4$, the band oscillation follows another pattern of type-II, where the band structures near the Fermi level contain a mixture of many bands that oscillate in phase. However, these bands are not coupled with each other. The band oscillation periods at low temperature below 30K for FeSe, $BaFe_2As_2$, $HgBa_2CuO_4$, $BiSrCaCuO_4$ are 169, 167, 68 and 55 fs, respectively. In previous experiments, a coherent $A_{1g}$ phonon of 185 fs in doped $BaFe_2As_2$ was observed,[46, 47] and the orbital-resolved energy oscillation with a period of 187 fs was directly observed using ultrafast experiments in strained FeSe grown on $SrTiO_3$ substrate.[43] Both are close to the predicted periods by our computation for the corresponding undoped parent materials of the two superconductors.

Furthermore, we find that the band oscillation behavior in $H_3S$ exhibits an important response to external pressure (Fig. S1). Without any pressure, the energy fluctuates in a random way. However, when a large pressure is applied, corresponding to the necessary experimental pressure to trigger superconductivity, the band oscillation becomes regular, e.g., like cuprates. Furthermore, the band frequency also increases with the pressure. At 500 GPa, which is commonly applied in experiment,[48] the oscillation period is about 23 fs.

Usually, random white noise fluctuations are expected by the thermal equilibrium movement. Thus, the electron-phonon interaction, even often being coupled in strongly correlated materials, was also thought to be random and thus the net effect may be wiped out with time evolution. Here, the very regular oscillation with time evolution caused by the thermodynamic fluctuation of the phonon is rather surprising, implying a nontrivial physics. We notice that certain special channel of electron-phonon interaction related to nonthermal physics has been reported in experiments for other non-superconductive materials previously,[49-51] where the band oscillation is stemmed from the selective electron-phonon coupling and a predominant phonon.

Here, for $WTe_2$, the selective electron-phonon coupling has an origin due to broken symmetry. The band states near the Fermi level are selectively coupled with $E$ phonon, which breaks the two-fold screw symmetry. The energy gap is thus protected by the screw symmetry. A spontaneous band gap is generated when the $E$ phonon oscillates. Other phonon modes are neglected due to very weak coupling with electrons near the Fermi level. In $WTe_2$, the bands near the Fermi level oscillates in a period half of the selected phonon mode from our computation. This implies that the phonons are the not direct physical factors for nonadiabatic excitation.

For other materials investigated by us, predominant phonon is the main reason for band oscillation. For FeSe and $BaFe_2As_2$, the band oscillation is driven by a $A_{1g}$ phonon mode,[43, 46, 47] where the states in a large energy range oscillate in the same period. As shown in Figs. 2ab, from our simulation the molecular dynamics of nuclei can be projected onto the oxide apical modes along the out-of-plane direction. The probability projected onto the apical mode with frequency of 17.8 THz (an $A_g$ mode)[12] is around 35%, and the probability projected onto the apical mode of 2.5 THz is around 20%. Our TDDFT simulation is in good accordance with previous experiments. Similarly, we find that in other cuprates the oxide apical phonon mode is also predominant. In experiments, the mode-selective periodic oscillation of energy has been observed in different cuprates.[52-55] Due to the challenge of time resolution, the oscillation at 17.8 THz was not observed directly in cuprates, but previous works still emphasize the apical oxygen mode at the frequency around 17 THz in both experiments and theory.[12, 56-58]

Furthermore, here we for the first time show in Fig. 2c that our TDDFT calculated band oscillation behavior exhibits a strong correlation with the superconductive behaviors in these superconductors with a broad distribution of $T_c$ and families. The calculated frequencies are in good accordance with the experimental ones (see Fig. 2d). The band oscillation magnitude increases with temperature in cuprates, while the frequency is fixed at different temperatures below a critical value of ~50 K. Above 60 K, we notice that the particular dynamic band structures of the undoped system, namely the crossing of the valence band and the conduction band, will be mixed with other parts of the electronic structure, and the oscillation disappears (as shown Fig. S2). However, we find that when $HgBa_2CuO_4$ is doped with holes, the oscillation is robust below 15% hole-doping. However, some other phonons will be activated and mix with oxygen apical phonon mode when it is heavily doped above 25%, which will cause the decoherence due to the complex interaction between phonons (Fig. S3). The trend upon doping is in good accordance with experimental phase diagram of cuprates.[59]

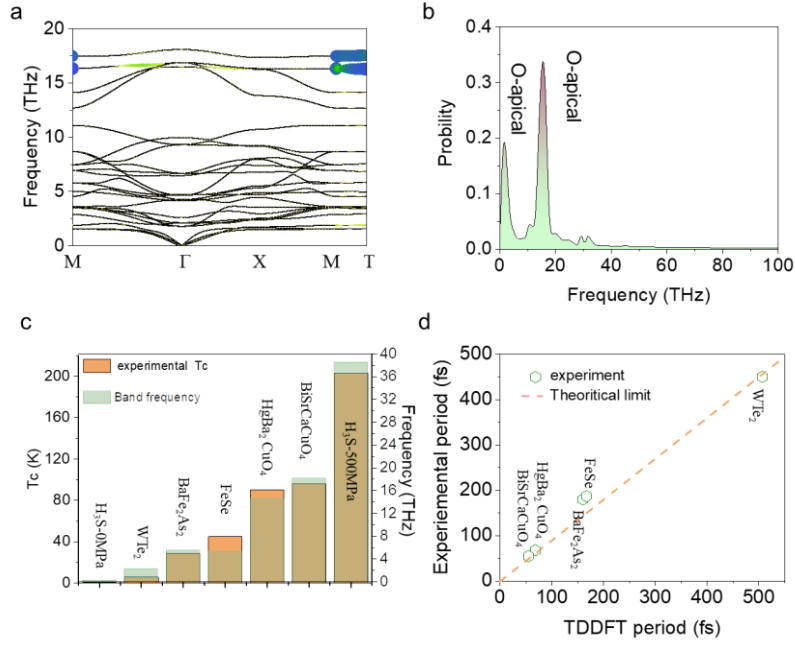

**Figure 2.** (a) Phonon band structure of HgBa$_2$CuO$_4$. Weight represents the phonon project on the apical mode. (b) Fourier transformation of O movement in TDDFT. (c) Comparison between superconductivity transition temperature in experiments and calculated band oscillation frequency. (d) Comparison between oscillation periods in ultrafast experiments and periods of band oscillation in TDDFT calculations.

As the band structure oscillates, the electrons are excited up and down between neighboring bands. This results in a coherent electron and hole excitation, beyond adiabatic process (as illustrated in Fig. 3a). In WTe$_2$, the electrons and holes switch in every half period. In iron-based compounds and cuprates, the electrons can be excited in even larger energy scale. The traditional picture of static Fermi level thus loses its physical importance in all materials here. If we monitor the charges in real space with time, they are found to be a time-dependent coherent charge density wave excited by phonon. At a given time, the local charge excitation is positive (holes) in some regions, and negative (electrons) in other alternating regions. Note that even for the same type of atoms, the local charge excitations can be opposite in sign. The wavelength of such charge density wave excitation is the same as the mode-selective phonon, which is usually longer than the unit cell size of these materials. Two examples are shown in Figs. 3bc, where the wave vectors of the charge density wave are ($2\pi$/a, $2\pi$/a, 0) and ($2\pi$/a, 0, 0) for HgBa$_2$CuO$_4$ and WTe$_2$, respectively. The induced charge fluctuation can be as large as 0.2 e per Cu atom in cuprates. The induced Coulomb interaction is with an energy scale as large as a few hundreds of meV, which is comparable to the "kink energy" from photoemission measurements,[60, 61] and the expected energy scale for superconductive paring as well. In addition, it can be observed that the charge excitation shows a d$_{x2-y2}$ symmetry in cuprates, and a *p* type symmetry in WTe$_2$.

To further understand our TDDFT simulation results, we build an effective minimum single-particle Hamiltonian as follows:

$$H = \begin{pmatrix} H_c(0) + B(t) & D_q(t) \\ D_q^\dagger(t) & H_v(q) + B(t) \end{pmatrix} \quad (1)$$

where $H_v$ and $H_c$ are the occupied and unoccupied sub-Hamiltonian without electron-phonon interaction. The electron-phonon interaction is treated as a time-dependent perturbation. $D_q(t) = \langle \psi_c | V_{ele-ion}(q,t) | \psi_v \rangle$ and $B(t) = \langle \psi_{c/v} | V_{ele-ion}(q,t) | \psi_{c/v} \rangle$ can be obtained by extending up to the first order according to the theory of degeneracy perturbation. Both $D_q(t)$ and $B(t)$ are time $t$ dependent.

We follow a similar algorithm with our TDDFT to obtain the solution of the time dependent Schrodinger's equation as $|\psi_k(t)\rangle = e^{-\frac{i}{\hbar}\int H_k dt}|\psi_0\rangle \approx \prod e^{-\frac{i}{\hbar}H_k \Delta t}|\psi_0\rangle$ [41, 62, 63]. We have

$$|\psi_k(t)\rangle = \begin{pmatrix} c_{11}(k,t) & c_{12}(k,t) \\ c_{21}(k,t) & c_{22}(k,t) \end{pmatrix} \begin{pmatrix} |\varphi_{vk}\rangle \\ |\varphi_{ck}\rangle \end{pmatrix} \quad (2)$$

Then the coefficients $c_{ij}(t)$ can be solved using Hamiltonian of equation (1). We have used the instantaneous eigenstates at $t = 0$ as the initial states of equation (2). This method is able to show the electron coherence and nonadiabatic excitation, which are beyond the matrix diagonalization. The obtained energy as a function of time is shown in Fig. 2f. $B(t)$ leads to Type-I electron phonon coupling, while $D(t)$ results in electron phonon coupling of Type-II. $B(t)$ and $D(t)$ can coexist in some materials.

$B(t)$ and $D_q(t)$ can be obtained by fitting the TDDFT simulation results or calculated in the perturbation method. In the case of WTe$_2$, a symmetry allowed Hamiltonian is $H_k = \tau v_{10} k_x \sigma_0 + \tau v_x k_x \sigma_x + v_y k_y \sigma_y$.[64] The energy gap is thus protected by the screw symmetry. Any phonons breaking the screw symmetry can induce a gap. The band structure of WTe$_2$ upon atomic perturbation can be found in Fig. S4. The perturbation calculations show the gap opening is sensitive to the atomic x-distortion of two neighbor W atoms moving relative to each other as shown in Fig. S5a. When we calculate weight breaking the screw symmetry of each possible phonon, we found only a few modes are predominant as shown in Fig. S5b. When electron-phonon coupling $\xi$ is considered, a time-dependent Hamiltonian is $H_k = \tau v_{10} k_x \sigma_0 + \tau v_x k_x \sigma_x + v_y k_y \sigma_y + \xi \sigma_z \sin(\omega t)$. Afer a unitory transfromtion, the electron phonon coupling is determined as $D(t) = \xi(0)\sin(\omega t)$, which couples two bands near the Fermi level. $\xi(0) = 0.1$ eV at 30K, and $\omega = 2\pi/T$ with $T$=420 fs.

In the case of cuprates compounds, $H_k = 2t(\cos(k_x) + \cos(k_y))$ based on a Cu-d$_{x2-y2}$ orbital.[65]

Because the apical phonon mode change the onsite energy, the electron phonon coupling is, $B(t) = \xi(0)\cos(\omega t)$ where $\xi(0) = 0.19$ eV at 10K, and $\omega = 2\pi/T$ with T=68 fs. Due to the symmetry of phonon and crystal, $B(t)$ is predominant over $D_q(t)$ in FeSe and cuprates near the Fermi level, while $D_q(t)$ is predominant in WTe$_2$. As shown in Fig. 1(f), the TDDFT simulation can be well reproduced by an effective model in Eq. 1. This implies that band oscillation is coherently driven by one or few selected phonon modes. According to Fig. 1, the calculated electron occupation is almost the same for an oscillating energy state. Electrons donot always occupy the lowest-energy bands. During half period of the band oscillation, the electron can temporarily reside on states with higher energy, creating free electrons and holes.

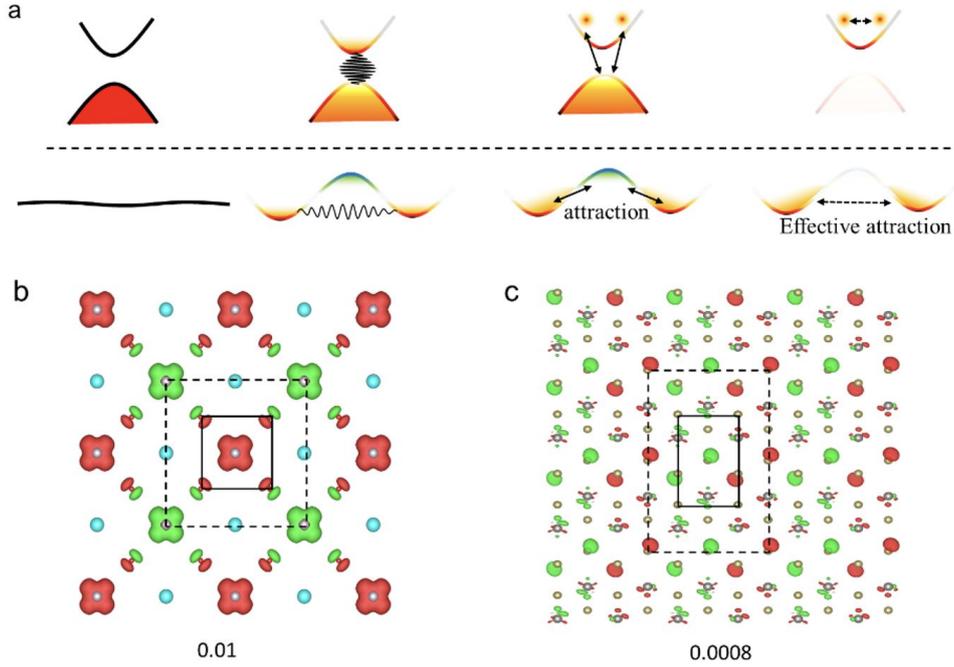

**Figure 3.** (a) Illusion of pairing mechanism under nonadiabatic excitation. Upper and lower panels are ultrafast excited charge in momentum space and real space. Wave line represents phonon. Dash and double arrows represent effective attractive interaction between two electrons (or holes). (b) DFT ultrafast excited charge oscillation in HgBa$_2$CuO$_4$ by an oxide apical phonon. (c) DFT ultrafast excited charge oscillation in monolayer WTe$_2$ by an *E* phonon mode. Red and green are local positive and negative charge excitations, respectively. The isosurface in a unit of e/Bohr$^3$ of the is labeled below the corresponding figures.

In experiments, 1*T'*-WTe$_2$ is an unconventional superconductor with a critical temperature of 1 ~ 6 K under low-density doping or pressure.[66-68] FeSe,[69] BaFe$_2$As$_2$,[70, 71] HgBa$_2$CuO$_4$,[72, 73] are also typical unconventional superconductors with critical temperatures of 45K, 29K and 93K. As shown in Fig. 2c, the TDDFT calculated oscillation frequencies show a strong correlation with $T_c$ in superconductors. Besides, H$_3$S is not a superconductor without pressure, while it becomes a high-$T_c$ superconductor at a high pressure,[74, 75] which correlates well with our simulated pressure response of band oscillation behavior in $H_3S$. This implies a nontrivial relevance between superconductivity and the band oscillation behavior unveiled by our simulation.

The DFT and TDDFT calculations include the electron-electron interaction in the method of mean-field theory through exchange-correlation functional. As an extended discussion, we include the electron-electron interaction in a model to find the possible reason for the correlation between band oscillation and critical temperatures.

$$H = \sum_{k\sigma}\varepsilon(\mathbf{k})a^\dagger_{\sigma k}a_{\sigma k} + \sum_{p\sigma'}\xi(p)c^\dagger_{\sigma' p}c_{\sigma' p} + \sum_{kp\sigma\sigma'}v(q)a^\dagger_{\sigma k+q}a_{\sigma k}c^\dagger_{\sigma' p-q}c_{\sigma' p} \tag{3}$$

Here, q indexes momentum, and $\varepsilon_0$ is the vacuum dielectric constant. $\alpha$ is screening constant. $a^\dagger_{\sigma k}(a_{\sigma k})$ is the creation (annihilation) operator for an electron with spin $\sigma$ and momentum $\mathbf{k}$, and $c^\dagger_{\sigma' p}(c_{\sigma' p})$ is the creation (annihilation) operator for a hole with spin $\sigma'$ and momentum $\mathbf{p}$. Then we performed a canonical transformation[76] as $\tilde{H} = e^{-S}He^S = H + [H,S] + \frac{1}{2}[[H,S],S] + ...$ with an ansatz $S = \sum_{kpq\sigma\sigma'} xv(q)a^\dagger_{\sigma k+q}a_{\sigma k}c^\dagger_{\sigma' p-q}c_{\sigma' p}$ and $S$ is anti-Hermitian. We arrived at (more details are in supporting information)

$$\tilde{H} = \sum_{k\sigma}\varepsilon(\mathbf{k})a^\dagger_{\sigma k}a_{\sigma k} + 2\sum_{kk_1q\sigma}V(q)(a^\dagger_{\sigma k+q}a_{\sigma k}a^\dagger_{\sigma_1 k_1-q}a_{\sigma_1 k_1}) + 2\sum_{kk_1q\sigma}V_1(q)(a^\dagger_{\sigma k+q}a_{\sigma k}a^\dagger_{\sigma' k_1+q}a_{\sigma' k_1}) \tag{4}$$

where the interaction $V(q) = \sum_{p\sigma'}xv(-q)v(q)(c^\dagger_{\sigma' p-q}c_{\sigma' p-q} - c^\dagger_{\sigma' p}c_{\sigma' p})$ can be negative, $V_1(q) = \sum_{p\sigma'}xv(q)v(q)(c^\dagger_{\sigma' p-q}c_{\sigma' p+q})$ and $x = 1/(\varepsilon(k+q)-\varepsilon(k)+\xi(p-q)-\xi(p))$, where $\varepsilon(k)$ and $\xi(p)$ are electron and hole energy dispersion, respectively. Thus, an effective attraction induced by electron excitation is possible.

If we perform the similar procedure of mean field approximation, (we show more details in SI) we arrive at a conclusion that the pseudogap and critical temperature of superconductors is proportional to the characteristic band frequency, and we obtained a relation of $k_B T_c = 1.13\hbar\omega_c e^{-1/N_0 V}$, where $N_0$ is the density of states near the Fermi level. This looks similar to BCS theory due to the shared mean field approximation. $\omega_c$ here is a characteristic frequency of charge or band oscillation. When the bands near the Fermi level are flat, $N_0 V$ is thus large and $e^{-1/N_0 V}$ is approximately close to 1. In this case, $T_c$ is proportional to $\omega_c$. This is in good accordance with our TDDFT simulation and experimental observation. In physics, the current proposal is different from previous BCS theory by mode-selective electron phonon coupling, nonadiabatical excitation and charge-wave-mediated interaction. More discussion about the difference between our proposal and the BCS theory is attached in the SI.

As shown in Fig. 3a, during the band oscillation, dynamic electrons or holes are nonadiabatically created. Electrons (holes) are attracted by holes (electrons) due to the interaction. An effective attraction between two electrons (or two holes) leads to a possible Cooper pairing. More details discussions can be seen in supporting information. The formula well explains why the band frequency is strongly correlated with the transition temperature of these superconductors.

To summary, we used TDDFT to simulate the electron dynamics of various main families of superconductors. We found the oscillation frequency of charges or bands here are strongly correlated with a broad range of superconductivity transition temperatures. The electron phonon coupling is mode-selective. Electron can be nonadiabatically excited, and keep its coherence. This implies the importance of coherence, mode-selective electron phonon coupling and nonadiabatic excitation for superconducting mechanism. We believe our numerical results will inspire the theoretical scientists to investigate the electron pairing mechanism in a new perspective.

**Acknowledgement**: This work was supported by the US Department of Energy and NSF.

[1] J.G. Bednorz, K.A. Müller, Possible high T c superconductivity in the Ba− La− Cu− O system, Zeitschrift für Physik B Condensed Matter, 64 (1986) 189-193.

[2] F. Giustino, Electron-phonon interactions from first principles, Reviews of Modern Physics, 89 (2017) 015003.

[3] M.P. Bircher, E. Liberatore, N.J. Browning, S. Brickel, C. Hofmann, A. Patoz, O.T. Unke, T. Zimmermann, M. Chergui, P. Hamm, Nonadiabatic effects in electronic and nuclear dynamics, Structural Dynamics, 4 (2017).

[4] A. Dienst, M.C. Hoffmann, D. Fausti, J.C. Petersen, S. Pyon, T. Takayama, H. Takagi, A. Cavalleri, Bi-directional ultrafast electric-field gating of interlayer charge transport in a cuprate superconductor, Nature Photonics, 5 (2011) 485-488.

[5] C. Giannetti, M. Capone, D. Fausti, M. Fabrizio, F. Parmigiani, D. Mihailovic, Ultrafast optical spectroscopy of strongly correlated materials and high-temperature superconductors: a non-equilibrium approach, Advances in Physics, 65 (2016) 58-238.

[6] C.L. Smallwood, J.P. Hinton, C. Jozwiak, W. Zhang, J.D. Koralek, H. Eisaki, D.-H. Lee, J. Orenstein, A. Lanzara, Tracking Cooper pairs in a cuprate superconductor by ultrafast angle-resolved photoemission, Science, 336 (2012) 1137-1139.

[7] R. Cortés, L. Rettig, Y. Yoshida, H. Eisaki, M. Wolf, U. Bovensiepen, Momentum-resolved ultrafast electron dynamics in superconducting $Bi_2Sr_2CaCu_2O_{8+\delta}$, Physical Review Letters, 107 (2011) 097002.

[8] A. De La Torre, D.M. Kennes, M. Claassen, S. Gerber, J.W. McIver, M.A. Sentef, Colloquium: Nonthermal pathways to ultrafast control in quantum materials, Reviews of Modern Physics, 93 (2021) 041002.

[9] M. Trigo, M.P. Dean, D.A. Reis, Ultrafast x-ray probes of dynamics in solids, arXiv preprint arXiv:2108.05456, DOI (2021).

[10] C.R. Hunt, D. Nicoletti, S. Kaiser, D. Pröpper, T. Loew, J. Porras, B. Keimer, A. Cavalleri, Dynamical decoherence of the light induced interlayer coupling in $YBa_2Cu_3O_{6+\delta}$, Physical Review B, 94 (2016) 224303.

[11] M. Budden, T. Gebert, M. Buzzi, G. Jotzu, E. Wang, T. Matsuyama, G. Meier, Y. Laplace, D. Pontiroli, M. Riccò, Evidence for metastable photo-induced superconductivity in K3C60, Nature Physics, 17 (2021) 611-618.

[12] R. Mankowsky, A. Subedi, M. Först, S.O. Mariager, M. Chollet, H. Lemke, J.S. Robinson, J.M. Glownia, M.P. Minitti, A. Frano, Nonlinear lattice dynamics as a basis for enhanced superconductivity in YBa2Cu3O6. 5, Nature, 516 (2014) 71-73.

[13] T. Suzuki, T. Someya, T. Hashimoto, S. Michimae, M. Watanabe, M. Fujisawa, T. Kanai, N. Ishii, J. Itatani, S. Kasahara, Photoinduced possible superconducting state with long-lived disproportionate band filling in FeSe, Communications Physics, 2 (2019) 115.

[14] R. Mankowsky, M. Fechner, M. Först, A. von Hoegen, J. Porras, T. Loew, G. Dakovski, M. Seaberg, S. Möller, G. Coslovich, Optically induced lattice deformations, electronic structure changes, and enhanced superconductivity in YBa2Cu3O6. 48, Structural Dynamics, 4 (2017).

[15] D. Fausti, R. Tobey, N. Dean, S. Kaiser, A. Dienst, M.C. Hoffmann, S. Pyon, T. Takayama, H. Takagi, A. Cavalleri, Light-induced superconductivity in a stripe-ordered cuprate, science, 331 (2011) 189-191.

[16] D.M. Kennes, E.Y. Wilner, D.R. Reichman, A.J. Millis, Transient superconductivity from electronic squeezing of optically pumped phonons, Nature Physics, 13 (2017) 479-483.

[17] A.S. Disa, T.F. Nova, A. Cavalleri, Engineering crystal structures with light, Nature Physics, 17

(2021) 1087-1092.

[18] D. Nevola, N. Zaki, J. Tranquada, W.-G. Yin, G. Gu, Q. Li, P. Johnson, Ultrafast Melting of Superconductivity in an Iron-Based Superconductor, Physical Review X, 13 (2023) 011001.

[19] C. Wang, X. Liu, Q. Chen, D. Chen, Y. Wang, S. Meng, Coherent-Phonon-Driven Intervalley Scattering and Rabi Oscillation in Multivalley 2D Materials, Physical Review Letters, 131 (2023) 066401.

[20] C. Lian, S.-J. Zhang, S.-Q. Hu, M.-X. Guan, S. Meng, Ultrafast charge ordering by self-amplified exciton–phonon dynamics in TiSe2, Nature communications, 11 (2020) 43.

[21] M.-X. Guan, X.-B. Liu, D.-Q. Chen, X.-Y. Li, Y.-P. Qi, Q. Yang, P.-W. You, S. Meng, Optical control of multistage phase transition via phonon coupling in MoTe 2, Physical Review Letters, 128 (2022) 015702.

[22] N. Punpongjareorn, X. He, Z. Tang, A.M. Guloy, D.-S. Yang, Ultrafast switching of valence and generation of coherent acoustic phonons in semiconducting rare-earth monosulfides, Applied Physics Letters, 111 (2017).

[23] J. Bang, Y. Sun, X.-Q. Liu, F. Gao, S. Zhang, Carrier-multiplication-induced structural change during ultrafast carrier relaxation and nonthermal phase transition in semiconductors, Physical Review Letters, 117 (2016) 126402.

[24] S. Sadasivam, M.K. Chan, P. Darancet, Theory of thermal relaxation of electrons in semiconductors, Physical review letters, 119 (2017) 136602.

[25] O.-J. Wacker, R. Kümmel, E. Gross, Time-dependent density-functional theory for superconductors, Physical review letters, 73 (1994) 2915.

[26] C. Lian, M. Guan, S. Hu, J. Zhang, S. Meng, Photoexcitation in solids: First-principles quantum simulations by real-time TDDFT, Advanced Theory and Simulations, 1 (2018) 1800055.

[27] P. You, D. Chen, C. Lian, C. Zhang, S. Meng, First-principles dynamics of photoexcited molecules and materials towards a quantum description, Wiley Interdisciplinary Reviews: Computational Molecular Science, 11 (2021) e1492.

[28] G. Kolesov, O. Grånäs, R. Hoyt, D. Vinichenko, E. Kaxiras, Real-time TD-DFT with classical ion dynamics: Methodology and applications, Journal of Chemical Theory and Computation, 12 (2016) 466-476.

[29] J.L. Alonso, X. Andrade, P. Echenique, F. Falceto, D. Prada-Gracia, A. Rubio, Efficient formalism for large-scale ab initio molecular dynamics based on time-dependent density functional theory, Physical review letters, 101 (2008) 096403.

[30] S. Meng, E. Kaxiras, Real-time, local basis-set implementation of time-dependent density functional theory for excited state dynamics simulations, The Journal of chemical physics, 129 (2008).

[31] A. Kolobov, P. Fons, J. Tominaga, M. Hase, Excitation-assisted disordering of GeTe and related solids with resonant bonding, The Journal of Physical Chemistry C, 118 (2014) 10248-10253.

[32] E. Runge, E.K. Gross, Density-functional theory for time-dependent systems, Physical review letters, 52 (1984) 997.

[33] M.A. Marques, E.K. Gross, Time-dependent density functional theory, Annu. Rev. Phys. Chem., 55 (2004) 427-455.

[34] E. Matsubara, S. Okada, T. Ichitsubo, T. Kawaguchi, A. Hirata, P. Guan, K. Tokuda, K. Tanimura, T. Matsunaga, M. Chen, Initial atomic motion immediately following femtosecond-laser excitation in phase-change materials, Physical review letters, 117 (2016) 135501.

[35] N.-K. Chen, X.-B. Li, J. Bang, X.-P. Wang, D. Han, D. West, S. Zhang, H.-B. Sun, Directional forces

by momentumless excitation and order-to-order transition in peierls-distorted solids: the case of GeTe, Physical Review Letters, 120 (2018) 185701.

[36] X.-B. Li, X. Liu, X. Liu, D. Han, Z. Zhang, X. Han, H.-B. Sun, S. Zhang, Role of electronic excitation in the amorphization of Ge-Sb-Te alloys, Physical review letters, 107 (2011) 015501.

[37] W.-H. Liu, J.-W. Luo, S.-S. Li, L.-W. Wang, The seeds and homogeneous nucleation of photoinduced nonthermal melting in semiconductors due to self-amplified local dynamic instability, Science advances, 8 (2022) eabn4430.

[38] M.C. Lucking, J. Bang, H. Terrones, Y.-Y. Sun, S. Zhang, Multivalency-induced band gap opening at $MoS_2$ edges, Chemistry of Materials, 27 (2015) 3326-3331.

[39] J. Ma, Z. Wang, L.-W. Wang, Interplay between plasmon and single-particle excitations in a metal nanocluster, Nature communications, 6 (2015) 10107.

[40] Z. Wang, S.-S. Li, L.-W. Wang, Efficient real-time time-dependent density functional theory method and its application to a collision of an ion with a 2D material, Physical Review Letters, 114 (2015) 063004.

[41] L.-W. Wang, Natural orbital branching scheme for time-dependent density functional theory nonadiabatic simulations, The Journal of Physical Chemistry A, 124 (2020) 9075-9087.

[42] W.H. Liu, Z. Wang, Z.H. Chen, J.W. Luo, S.S. Li, L.W. Wang, Algorithm advances and applications of time-dependent first-principles simulations for ultrafast dynamics, Wiley Interdisciplinary Reviews: Computational Molecular Science, 12 (2022) e1577.

[43] S. Gerber, S.-L. Yang, D. Zhu, H. Soifer, J. Sobota, S. Rebec, J. Lee, T. Jia, B. Moritz, C. Jia, Femtosecond electron-phonon lock-in by photoemission and x-ray free-electron laser, Science, 357 (2017) 71-75.

[44] P. Hein, S. Jauernik, H. Erk, L. Yang, Y. Qi, Y. Sun, C. Felser, M. Bauer, Mode-resolved reciprocal space mapping of electron-phonon interaction in the Weyl semimetal candidate $T_d$-$WTe_2$, Nature communications, 11 (2020) 2613.

[45] D. Soranzio, M. Savoini, P. Beaud, F. Cilento, L. Boie, J. Dössegger, V. Ovuka, S. Houver, M. Sander, S. Zerdane, Strong modulation of carrier effective mass in $WTe_2$ via coherent lattice manipulation, npj 2D Materials and Applications, 6 (2022) 71.

[46] L. Yang, G. Rohde, T. Rohwer, A. Stange, K. Hanff, C. Sohrt, L. Rettig, R. Cortés, F. Chen, D. Feng, Ultrafast modulation of the chemical potential in $BaFe_2As_2$ by coherent phonons, Physical Review Letters, 112 (2014) 207001.

[47] K.W. Kim, A. Pashkin, H. Schäfer, M. Beyer, M. Porer, T. Wolf, C. Bernhard, J. Demsar, R. Huber, A. Leitenstorfer, Ultrafast transient generation of spin-density-wave order in the normal state of $BaFe_2As_2$ driven by coherent lattice vibrations, Nature materials, 11 (2012) 497-501.

[48] A.P. Durajski, R. Szczęśniak, First-principles study of superconducting hydrogen sulfide at pressure up to 500 GPa, Scientific Reports, 7 (2017) 4473.

[49] H. Hübener, U. De Giovannini, A. Rubio, Phonon driven Floquet matter, Nano Letters, 18 (2018) 1535-1542.

[50] Z. Song, L.-W. Wang, Electron-phonon coupling induced intrinsic Floquet electronic structure, npj Quantum Materials, 5 (2020) 77.

[51] S. Zhou, C. Bao, B. Fan, H. Zhou, Q. Gao, H. Zhong, T. Lin, H. Liu, P. Yu, P. Tang, Pseudospin-selective Floquet band engineering in black phosphorus, Nature, 614 (2023) 75-80.

[52] M. Forst, R. Mankowsky, A. Cavalleri, Mode-selective control of the crystal lattice, Accounts of chemical research, 48 (2015) 380-387.

[53] T. Dong, S.J. Zhang, N.L. Wang, Recent development of ultrafast optical characterizations for

quantum materials, Advanced Materials, 35 (2023) 2110068.

[54] A. Cavalleri, Photo-induced superconductivity, Contemporary Physics, 59 (2018) 31-46.

[55] A. von Hoegen, M. Fechner, M. Först, N. Taherian, E. Rowe, A. Ribak, J. Porras, B. Keimer, M. Michael, E. Demler, Amplification of superconducting fluctuations in driven $YBa_2Cu_3O_{6+x}$, Physical Review X, 12 (2022) 031008.

[56] S. Johnston, F. Vernay, B. Moritz, Z.-X. Shen, N. Nagaosa, J. Zaanen, T. Devereaux, Systematic study of electron-phonon coupling to oxygen modes across the cuprates, Physical Review B, 82 (2010) 064513.

[57] S.-L. Yang, J. Sobota, Y. He, D. Leuenberger, H. Soifer, H. Eisaki, P. Kirchmann, Z.-X. Shen, Mode-selective coupling of coherent phonons to the Bi2212 electronic band structure, Physical review letters, 122 (2019) 176403.

[58] S. Kaiser, C.R. Hunt, D. Nicoletti, W. Hu, I. Gierz, H. Liu, M. Le Tacon, T. Loew, D. Haug, B. Keimer, Optically induced coherent transport far above $T_c$ in underdoped $YBa_2Cu_3O_{6+\delta}$, Physical Review B, 89 (2014) 184516.

[59] N. Armitage, P. Fournier, R. Greene, Progress and perspectives on electron-doped cuprates, Reviews of Modern Physics, 82 (2010) 2421.

[60] A. Lanzara, P. Bogdanov, X. Zhou, S. Kellar, D. Feng, E. Lu, T. Yoshida, H. Eisaki, A. Fujimori, K. Kishio, Evidence for ubiquitous strong electron–phonon coupling in high-temperature superconductors, Nature, 412 (2001) 510-514.

[61] Z.-X. Shen, A. Lanzara, S. Ishihara, N. Nagaosa, Role of the electron-phonon interaction in the strongly correlated cuprate superconductors, Philosophical magazine B, 82 (2002) 1349-1368.

[62] J.J. Goings, P.J. Lestrange, X. Li, Real-time time-dependent electronic structure theory, Wiley Interdisciplinary Reviews: Computational Molecular Science, 8 (2018) e1341.

[63] A. Castro, M.A. Marques, A. Rubio, Propagators for the time-dependent Kohn–Sham equations, The Journal of chemical physics, 121 (2004) 3425-3433.

[64] L. Muechler, A. Alexandradinata, T. Neupert, R. Car, Topological nonsymmorphic metals from band inversion, Physical Review X, 6 (2016) 041069.

[65] M. Hirayama, Y. Yamaji, T. Misawa, M. Imada, Ab initio effective Hamiltonians for cuprate superconductors, Physical Review B, 98 (2018) 134501.

[66] W. Yang, C.-J. Mo, S.-B. Fu, Y. Yang, F.-W. Zheng, X.-H. Wang, Y.-A. Liu, N. Hao, P. Zhang, Soft-Mode-Phonon-Mediated Unconventional Superconductivity in Monolayer $1T'-WTe_2$, Physical Review Letters, 125 (2020) 237006.

[67] X.-C. Pan, X. Chen, H. Liu, Y. Feng, Z. Wei, Y. Zhou, Z. Chi, L. Pi, F. Yen, F. Song, Pressure-driven dome-shaped superconductivity and electronic structural evolution in tungsten ditelluride, Nature communications, 6 (2015) 7805.

[68] V. Fatemi, S. Wu, Y. Cao, L. Bretheau, Q.D. Gibson, K. Watanabe, T. Taniguchi, R.J. Cava, P. Jarillo-Herrero, Electrically tunable low-density superconductivity in a monolayer topological insulator, Science, 362 (2018) 926-929.

[69] F.-C. Hsu, J.-Y. Luo, K.-W. Yeh, T.-K. Chen, T.-W. Huang, P.M. Wu, Y.-C. Lee, Y.-L. Huang, Y.-Y. Chu, D.-C. Yan, Superconductivity in the PbO-type structure α-FeSe, Proceedings of the National Academy of Sciences, 105 (2008) 14262-14264.

[70] D. Mandrus, A.S. Sefat, M.A. McGuire, B.C. Sales, Materials chemistry of $BaFe_2As_2$: a model platform for unconventional superconductivity, Chemistry of Materials, 22 (2010) 715-723.

[71] P.L. Alireza, Y.C. Ko, J. Gillett, C.M. Petrone, J.M. Cole, G.G. Lonzarich, S.E. Sebastian,

Superconductivity up to 29 K in SrFe2As2 and BaFe2As2 at high pressures, Journal of Physics: Condensed Matter, 21 (2008) 012208.

[72] S. Putilin, E. Antipov, O. Chmaissem, M. Marezio, Superconductivity at 94 K in HgBa2CuO4+ δ, Nature, 362 (1993) 226-228.

[73] S. Sadewasser, J. Schilling, J. Wagner, O. Chmaissem, J. Jorgensen, D. Hinks, B. Dabrowski, Relaxation effects in the transition temperature of superconducting HgBa 2 CuO 4+ δ, Physical Review B, 60 (1999) 9827.

[74] D. Duan, Y. Liu, F. Tian, D. Li, X. Huang, Z. Zhao, H. Yu, B. Liu, W. Tian, T. Cui, Pressure-induced metallization of dense (H2S) 2H2 with high-T c superconductivity, Scientific reports, 4 (2014) 6968.

[75] F. Capitani, B. Langerome, J.-B. Brubach, P. Roy, A. Drozdov, M. Eremets, E. Nicol, J. Carbotte, T. Timusk, Spectroscopic evidence of a new energy scale for superconductivity in H 3 S, Nature physics, 13 (2017) 859-863.

[76] I. Lang, Y.A. Firsov, Kinetic theory of semiconductors with low mobility, Sov. Phys. JETP, 16 (1963) 1301.